\documentclass[useAMS,usenatbib]{mn2e}
\usepackage{epsfig,graphicx,lscape}
\title[Fe X lines in SERTS spectra]{Emission lines of Fe XI in the 257--407\,\AA\ wavelength region observed in solar spectra from EIS/Hinode and SERTS}

\author[F. P. Keenan et al.]{F. P. Keenan\thanks{E-mail:
F.Keenan@qub.ac.uk},$^{1}$ R. O. Milligan,$^{2}$ D. B. Jess,$^{1}$ K. M. Aggarwal,$^{1}$ M. Mathioudakis,$^{1}$
\newauthor
R. J. Thomas,$^{2}$  J. W. Brosius$^{2,3}$ and J. M. Davila$^{2}$
\\
$^{1}$Astrophysics Research Centre, School of Mathematics and Physics, Queen's University, Belfast BT7 1NN
\\
$^{2}$Laboratory for Solar Physics, Code 671, Heliophysics Science Division,
NASA Goddard Space Flight Center, Greenbelt, 
\\
MD 20771, USA
\\
$^{3}$Department of Physics, The Catholic University of
America, Washington, DC 20064, USA
}

\begin{document}

\date{Received 2009 xxxx}

\pagerange{\pageref{firstpage}--\pageref{lastpage}} \pubyear{}

\maketitle

\label{firstpage}

\begin{abstract}
Theoretical emission-line ratios involving Fe\,{\sc xi} transitions in the 257--407\,\AA\ wavelength range are derived using fully relativistic calculations of radiative rates and electron impact
excitation cross sections. These are subsequently compared with both long wavelength channel Extreme-Ultraviolet 
Imaging Spectrometer (EIS) spectra from the {\em Hinode} satellite (covering 245--291\,\AA), and first-order
observations ($\sim$\,235--449\,\AA) obtained by the Solar Extreme-ultraviolet Research Telescope and Spectrograph (SERTS).  The 266.39, 266.60 and 276.36\,\AA\ lines of Fe\,{\sc xi} are detected in two EIS spectra, confirming earlier identifications of these features, and 276.36\,\AA\ is found to provide an electron density (N$_{e}$) diagnostic when ratioed against the 257.55\,\AA\ transition. Agreement between theory and observation is found to be generally good for the 
SERTS data sets, with discrepancies normally being due to known line blends, while the 
257.55\,\AA\ feature is detected for the first time in SERTS spectra. 
The most useful Fe\,{\sc xi} electron density diagnostic is found to be the 308.54/352.67 intensity ratio, which varies by a factor of 
8.4 between N$_{e}$ = 10$^{8}$ and 10$^{11}$ cm$^{-3}$, while showing little temperature
sensitivity. However, the 349.04/352.67 ratio potentially provides a superior diagnostic, as it involves lines which are closer in wavelength, and varies by a factor of 14.7 between  N$_{e}$ = 10$^{8}$ and 10$^{11}$ cm$^{-3}$.
Unfortunately, the 349.04\,\AA\ line is relatively weak, and also blended with the second-order
Fe\,{\sc x} 174.52\,\AA\ feature, unless the first-order instrument response is enhanced.

\end{abstract}


\section{Introduction}

Emission lines arising from transitions in sulphur-like Fe\,{\sc xi} have been detected in the spectra of 
a range of astrophysical sources, including the Sun (Dere 1978) and other cool stars (Laming \& 
Drake 1999), as well as active galaxies (Puchnarewicz et al. 1998). 
The usefulness of these lines as electron density diagnostics for the emitting plasma 
was first shown by Kastner \& Mason (1978), who employed the atomic physics calculations of 
Mason (1975) to generate theoretical emission-line ratios for this ion.
Since then, several authors have calculated atomic data and theoretical line ratios for Fe\,{\sc xi}, 
particularly for applications to solar spectra (see Bhatia et al. 2002 and references therein).

Most work on the solar spectrum of Fe\,{\sc xi} has focused on the $\sim$\,180--200\,\AA\ 
wavelength range, containing 
3s$^{2}$3p$^{4}$--3s$^{2}$3p$^{3}$3d 
transitions (see, for example, Keenan et al. 2005). 
This is understandable, as this wavelength region 
contains numerous strong Fe\,{\sc xi} lines (including the most intense in the solar spectrum), many of which
provide good electron density diagnostics.
However, the $\sim$\,257--407\,\AA\ region also
contains several Fe\,{\sc xi} lines, primarily arising from 
3s$^{2}$3p$^{4}$--3s3p$^{5}$ transitions. 
In this paper we use the most recent atomic physics calculations for 
Fe\,{\sc xi} to analyse solar spectra covering this wavelength region.
Specifically, we investigate observational data sets from two missions, namely the 
Extreme-Ultraviolet Imaging Spectrometer (EIS) on board the {\em Hinode} satellite, and the Solar 
Extreme-ultraviolet Research Telescope and Spectrograph (SERTS) sounding rocket experiments.
The EIS long wavelength channel spectral region is 245--291\,\AA\ (Culhane et al. 2007), compared to $\sim$\,235--449\,\AA\ 
for the SERTS first-order bandpass (Neupert et al. 1992), and both have spectral resolutions of about
0.06\,\AA\ (FWHM). Although SERTS has a larger wavelength coverage, in principle allowing 
more lines from the same species to be detected, it has a limited dynamic range as most spectra 
have been recorded on photographic film. By contrast, EIS spectra are obtained with CCDs, and hence can reliably detect
weaker lines.
The combination of wavelength coverage and dynamic range provided by SERTS and EIS allows us to undertake a 
detailed assessment of the Fe\,{\sc xi} emission-line spectrum in the 257--407\,\AA\
wavelength region.

\section{Theoretical line ratios}

The model ion adopted for Fe\,{\sc xi} has been discussed in detail by Keenan et al. (2005).
Briefly, it consisted of the 24
energetically lowest LS states, yielding a total of 48 fine-structure levels.
Experimental energy levels, where available, were obtained from
Shirai et al. (1990) and Jup\'{e}n et al. (1993), with the 
theoretical results of Aggarwal \& Keenan (2003a)
being used for the remainder. Electron impact rates were taken from Aggarwal \& Keenan (2003b), 
while for Einstein A-coefficients the data of 
Aggarwal \& Keenan
(2003a) were employed. Finally, under solar plasma conditions proton impact 
excitation is only important for
transitions within the 3s$^{2}$3p$^{4}$ $^{3}$P ground term, 
and we have used the calculations of
Landman (1980) for this atomic process.

Using the model ion outlined above, in conjunction
with an updated version of the statistical equilibrium
code of Dufton (1977), relative Fe\,{\sc xi} emission-line 
strengths were calculated as a function of both electron temperature
(T$_{e}$) and density (N$_{e}$).
Details of the procedures involved and approximations made
may be found in
Dufton (1977) and Dufton et al.
(1978). Given errors in the adopted atomic data of typically $\pm$10 per cent (see the references above), we would expect the theoretical ratios to be accurate to better than $\pm$20 per cent. 

\section{Observational data}

The solar spectra analysed in the present paper are from two instruments, namely the EIS 
on board the {\em Hinode}
satellite, and several SERTS rocket flights. Our EIS data sets are those of an active region and limb area 
originally presented and discussed by Brown et al. (2008), to which the reader is referred
for details of the EIS instrument and the data reduction and 
flux calibration procedures. These two features, denoted AR1 and Limb by Brown et al., show the strongest Fe\,{\sc xi}
emission lines of the solar regions considered by these authors.
The spectra are from the EIS long wavelength channel and cover  
245--291\,\AA\ at a resolution of about 55\,m\AA\ (FWHM).

Our SERTS spectra are those of several quiet and active regions, plus an off-limb area,
obtained during rocket flights in 1989 (denoted SERTS89), 
1991 (SERTS91), 1993 (SERTS93) and 1997 (SERTS97). Details of these observations, including their reduction and
flux
calibration, may be found in Thomas \& Neupert (1994) for SERTS89, Brosius et al. (1996) for 
SERTS91 and SERTS93, and Brosius et al. (2000) for SERTS97.
Most of these data sets cover the 235-449\,\AA\ (SERTS89) or 231--445\,\AA\ (SERTS91 and SERTS93) wavelength regions in first-order, at resolutions of between 50--80\,m\AA, with spectra recorded on photographic film. 
However, the intensity calibrations for the SERTS91 and SERTS93 spectra are highly uncertain
for wavelengths shorter than 274\,\AA. By contrast,
the SERTS97 spectrum spans 299--353\,\AA\ in first-order, at a resolution of 115\,m\AA, and was recorded on a CCD. We note that the version of SERTS flown in 1989 carried a standard gold-coated toroidal diffraction grating, while those in 1991, 1993 and 1997 incorporated a multilayered-coated grating which enhanced the instrument sensitivity in the first-order wavelength range, by factors of up to 9. In addition, the absolute 
intensity calibration of the SERTS89 spectrum has been re-evaluated since the publication of Thomas \& 
Neupert, as a result of which emission-line intensities are uniformly a factor of 1.24 larger than those quoted in that
paper (see, for example, Keenan et al. 2008 and references therein). 

\section{Results and discussion}

We have searched for Fe\,{\sc xi} emission lines in the EIS and SERTS spectra 
using the detections of Brown et al. (2008) and Thomas \& Neupert (1994), 
supplemented with those
from other sources, including
the National Institute of Standards and Technology (NIST) data base,\footnote{http://physics.nist.gov/PhysRefData/} 
the latest version (V6.0) of the {\sc chianti}
data base (Dere et al. 1997, 2009), and the Atomic Line List of 
van Hoof.\footnote{http://www.pa.uky.edu/$\sim$peter/atomic/} 
 In Table 1 we list the Fe\,{\sc xi}
transitions identified in the data sets, along 
with their measured
wavelengths. We also provide a note on each line, including
whether it is detected in both EIS and SERTS spectra, and if there are possible blending features.

\begin{table*}
 \begin{minipage}{180mm}
  \caption{Fe\,{\sc xi} line identifications in the EIS and SERTS spectra.}
  \begin{tabular}{cll}
  \hline
Wavelength (\AA)    &   Transition  & Note
\\
\hline
257.55 & 3s$^{2}$3p$^{4}$ $^{3}$P$_{2}$--3s$^{2}$3p$^{3}$($^{4}$S)3d $^{5}$D$_{3}$ &  Detected by both EIS and SERTS.
\\
257.77 & 3s$^{2}$3p$^{4}$ $^{3}$P$_{2}$--3s$^{2}$3p$^{3}$($^{4}$S)3d $^{5}$D$_{2}$ &  Detected by EIS. 
\\
266.39 & 3s$^{2}$3p$^{4}$ $^{3}$P$_{1}$--3s$^{2}$3p$^{3}$($^{4}$S)3d $^{5}$D$_{1}$ &  Detected by EIS. 
\\
266.60 & 3s$^{2}$3p$^{4}$ $^{3}$P$_{1}$--3s$^{2}$3p$^{3}$($^{4}$S)3d $^{5}$D$_{0}$ &  Detected by EIS. 
\\
276.36 & 3s$^{2}$3p$^{4}$ $^{3}$P$_{2}$--3s3p$^{5}$ $^{1}$P$_{1}$  &  Detected by EIS.
\\
308.54 & 3s$^{2}$3p$^{4}$ $^{1}$D$_{2}$--3s3p$^{5}$ $^{1}$P$_{1}$ &  Detected by SERTS. Blended with Fe\,{\sc vi} (Brosius et al. 1998). 
\\
341.11 & 3s$^{2}$3p$^{4}$ $^{3}$P$_{2}$--3s3p$^{5}$ $^{3}$P$_{1}$ &  Detected by SERTS.
\\
349.04 & 3s$^{2}$3p$^{4}$ $^{3}$P$_{1}$--3s3p$^{5}$ $^{3}$P$_{0}$ &  Detected by SERTS. Blended with 2nd-order Fe\,{\sc x} 174.52\,\AA\ (Thomas \& Neupert 1994).
\\
352.67 & 3s$^{2}$3p$^{4}$ $^{3}$P$_{2}$--3s3p$^{5}$ $^{3}$P$_{2}$ &  Detected by SERTS.
\\
356.53 & 3s$^{2}$3p$^{4}$ $^{3}$P$_{1}$--3s3p$^{5}$ $^{3}$P$_{1}$ &  Detected by SERTS.
\\
358.67 & 3s$^{2}$3p$^{4}$ $^{3}$P$_{0}$--3s3p$^{5}$ $^{3}$P$_{1}$ &  Detected by SERTS. Blended with several lines (Young et al. 1998). 
\\
369.16 & 3s$^{2}$3p$^{4}$ $^{3}$P$_{1}$--3s3p$^{5}$ $^{3}$P$_{2}$ &  Detected by SERTS.
\\
406.79 & 3s$^{2}$3p$^{4}$ $^{1}$D$_{2}$--3s3p$^{5}$ $^{3}$P$_{2}$ & Detected by SERTS. Identified in spectrum by Brickhouse et al. (1995).
\\
\hline
\end{tabular}

\end{minipage} 
\end{table*}

Intensities and 
line widths (FWHM) of the Fe\,{\sc xi} features are given in Tables 2 and
3--9  for the EIS and SERTS data sets, respectively, 
along with the associated 1$\sigma$ errors. In the case of EIS, we have remeasured the spectra considered by Brown et al. (2008) 
using a standard least-squares fitting routine in SSWIDL assuming Gaussian profiles and a constant background. For 
SERTS the measurements were made with 
modified versions of the Gaussian fitting routines originally employed by Thomas \& Neupert (1994),
as discussed by Keenan et al. (2007).
As a consequence, the EIS and
SERTS line intensities, FWHM values and their errors listed in Tables 2--9
are somewhat different from those originally
reported in Brown et al., Thomas \& Neupert and Brosius et al. (1996, 2000). However,
in most directly comparable cases, the measured intensity and width 
values usually differ only slightly from those previously obtained. 
An exception is the 349.04\,\AA\ line in the SERTS spectra, where the features around 349\,\AA\ have been fitted with a triple Gaussian as opposed to the double Gaussian adopted by Thomas \& Neupert and Brosius et al. This was to allow for the presence of the 
weak Fe\,{\sc x} 349.30\,\AA\ transition, instead of only Mg\,{\sc vi} 349.16\,\AA.

\begin{figure*}
\epsfig{file=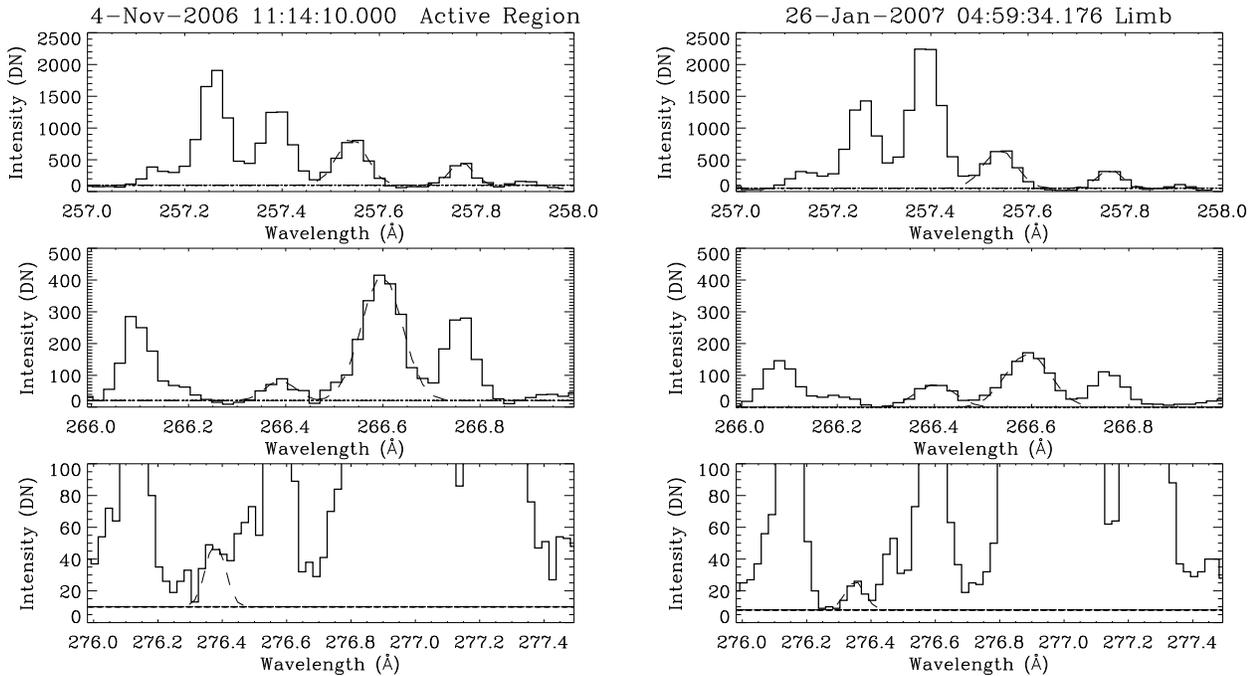,angle=90,width=16.5cm}
\caption{Portions of the EIS active region and limb spectra containing the Fe\,{\sc xi} 257.55, 257.77, 
266.39, 266.60 and 276.36\,\AA\ emission lines. The profile fits to these features are shown by dashed lines.} 
\end{figure*}

\begin{table}
  \caption{Fe\,{\sc xi} line intensities and widths from the EIS spectra.}
  \begin{tabular}{lccc}
  \hline
Feature$^{a}$ & Wavelength    &   Intensity & Line width 
\\
& (\AA) & (erg\,cm$^{-2}$\,s$^{-1}$\,sr$^{-1}$) & (m\AA) 
\\
\hline
Active region & 257.55 & 335.0 $\pm$ 31.1 & 73 $\pm$ 5 
\\
& 257.77 & 167.0 $\pm$ 24.2 & 55 $\pm$ 8 
\\
& 266.39 & 17.0 $\pm$ 3.4 & 70 $\pm$ 4 
\\
& 266.60 & 107.0 $\pm$ 9.1 & 90 $\pm$ 4 
\\
& 276.36 & 16.0 $\pm$ 7.7 & 82 $\pm$ 10 
\\
Limb & 257.55 & 278 $\pm$ 23.9 & 77 $\pm$ 5 
\\
& 257.77 & 133.0 $\pm$ 14.2 & 59 $\pm$ 9 
\\
& 266.39 & 19.0 $\pm$ 3.3 & 96 $\pm$ 14 
\\
& 266.60 & 45.0 $\pm$ 6.8 & 100 $\pm$ 9 
\\
& 276.36 & 6.0 $\pm$ 2.5 & 85 $\pm$ 10 
\\
\hline
\end{tabular}

$^{a}$The active region and limb features are those denoted AR1 and Limb, respectively, by Brown et al. (2008).

\end{table}

\begin{table}
  \caption{Fe\,{\sc xi} line intensities and widths from the SERTS 1989 active region spectrum (SERTS89--AR).}
  \begin{tabular}{ccc}
  \hline
Wavelength    &   Intensity & Line width 
\\
(\AA) & (erg\,cm$^{-2}$\,s$^{-1}$\,sr$^{-1}$) & (m\AA) 
\\
\hline
257.55 & 35.5 $\pm$15.1 & 96 $\pm$ 10 
\\
308.54$^{a}$ & 106.5 $\pm$ 20.3 & 122 $\pm$ 12
\\
341.11 & 46.1 $\pm$ 6.9 & 93 $\pm$ 8 
\\
349.04 & 26.2 $\pm$ 11.9 & 105 $\pm$ 39 
\\
352.67 & 161.6 $\pm$ 18.8 & 95 $\pm$ 6 
\\
356.53 & 22.1 $\pm$ 4.8 & 92 $\pm$ 17 
\\
358.67 & 88.7 $\pm$ 11.4 & 144 $\pm$ 9 
\\
369.16 & 47.1 $\pm$ 6.3 & 102 $\pm$ 5 
\\
406.79 & 3.6 $\pm$ 1.4 & 57 $\pm$ 13 
\\
\hline
\end{tabular}

$^{a}$Line measured at wavelength of 308.58\,\AA\ in this spectrum (see Section 4.2 for details).
\end{table}

\begin{table}
  \caption{Fe\,{\sc xi} line intensities and widths from the SERTS 1991 active region spectrum (SERTS91--AR).}
  \begin{tabular}{ccc}
  \hline
Wavelength    &   Intensity & Line width 
\\
(\AA) & (erg\,cm$^{-2}$\,s$^{-1}$\,sr$^{-1}$) & (m\AA) 
\\
\hline
308.54 & 5.9 $\pm$ 1.7 & 40 $\pm$ 10
\\
341.11 & 20.3 $\pm$ 3.7 & 56 $\pm$ 12 
\\
352.67 & 36.8 $\pm$ 5.6 & 57 $\pm$ 7 
\\
358.67 & 19.4 $\pm$ 4.9 & 64 $\pm$ 20 
\\
369.16 & 13.4 $\pm$ 3.6 & 69 $\pm$ 10 
\\
\hline
\end{tabular}
\end{table}

\begin{table}
  \caption{Fe\,{\sc xi} line intensities and widths from the SERTS 1991 quiet Sun spectrum (SERTS91--QS).}
  \begin{tabular}{ccc}
  \hline
Wavelength    &   Intensity & Line width 
\\
(\AA) & (erg\,cm$^{-2}$\,s$^{-1}$\,sr$^{-1}$) & (m\AA) 
\\
\hline
308.54 & 19.2 $\pm$ 5.7 & 38 $\pm$ 5
\\
341.11 & 58.6 $\pm$ 10.3 & 81 $\pm$ 14 
\\
352.67 & 112.6 $\pm$ 16.1 & 58 $\pm$ 10 
\\
356.53 & 17.2 $\pm$ 7.8 & 45 $\pm$ 15 
\\
358.67 & 14.2 $\pm$ 8.1 & 41 $\pm$ 16 
\\
369.16 & 44.1$\pm$ 12.3 & 70 $\pm$ 22 
\\
\hline
\end{tabular}

\end{table}

\begin{table}
  \caption{Fe\,{\sc xi} line intensities and widths from the SERTS 1991 off-limb spectrum (SERTS91--OL).}
  \begin{tabular}{ccc}
  \hline
Wavelength    &   Intensity & Line width 
\\
(\AA) & (erg\,cm$^{-2}$\,s$^{-1}$\,sr$^{-1}$) & (m\AA) 
\\
\hline
308.54 & 42.8 $\pm$ 13.1 & 43 $\pm$ 10
\\
341.11 & 99.7 $\pm$ 15.1 & 94 $\pm$ 12 
\\
352.67 & 240.8 $\pm$ 29.1 & 72 $\pm$ 8
\\
358.67 & 36.3 $\pm$ 18.2 & 37 $\pm$ 10 
\\
369.16 & 59.9 $\pm$ 8.3 & 109 $\pm$ 30 
\\
\hline
\end{tabular}

\end{table}

\begin{table}
  \caption{Fe\,{\sc xi} line intensities and widths from the SERTS 1993 active region spectrum (SERTS93--AR).}
  \begin{tabular}{ccc}
  \hline
Wavelength    &   Intensity & Line width 
\\
(\AA) & (erg\,cm$^{-2}$\,s$^{-1}$\,sr$^{-1}$) & (m\AA) 
\\
\hline
308.54 & 28.7 $\pm$ 11.6 & 45 $\pm$ 15
\\
341.11 & 71.1 $\pm$ 10.1 & 62 $\pm$ 9 
\\
352.67 & 178.1 $\pm$ 20.5 & 57 $\pm$ 9 
\\
356.53 & 19.3 $\pm$ 8.5 & 42 $\pm$ 25 
\\
358.67 & 61.1 $\pm$ 11.5 & 100 $\pm$ 17 
\\
369.16 & 59.8 $\pm$ 13.4 & 103 $\pm$ 22 
\\
\hline
\end{tabular}

\end{table}

\begin{table}
  \caption{Fe\,{\sc xi} line intensities and widths from the SERTS 1993 quiet Sun spectrum (SERTS93--QS).}
  \begin{tabular}{ccc}
  \hline
Wavelength    &   Intensity & Line width 
\\
(\AA) & (erg\,cm$^{-2}$\,s$^{-1}$\,sr$^{-1}$) & (m\AA) 
\\
\hline
308.54 & 6.6 $\pm$ 3.2 & 32 $\pm$ 12
\\
341.11 & 13.5 $\pm$ 2.8 & 50 $\pm$ 10 
\\
352.67 & 48.3 $\pm$ 6.5 & 63 $\pm$ 9 
\\
358.67 & 10.3 $\pm$ 2.9 & 83 $\pm$ 16 
\\
369.16 & 13.2 $\pm$ 3.4 & 115 $\pm$ 23 
\\
\hline
\end{tabular}

\end{table}

\begin{table}
  \caption{Fe\,{\sc xi} line intensities and widths from the SERTS 1997 active region spectrum (SERTS97--AR).}
  \begin{tabular}{ccc}
  \hline
Wavelength    &   Intensity & Line width 
\\
(\AA) & (erg\,cm$^{-2}$\,s$^{-1}$\,sr$^{-1}$) & (m\AA) 
\\
\hline
308.54 & 33.9 $\pm$ 9.8 & 142 $\pm$ 8
\\
341.11 & 37.8 $\pm$ 6.7 & 149 $\pm$ 9 
\\
349.04 & 12.5 $\pm$ 4.2 & 103 $\pm$ 9 
\\
352.67 & 136.1 $\pm$ 22.1 & 135 $\pm$ 9 
\\
\hline
\end{tabular}
\end{table}

\subsection{EIS spectra}

Although {\sc chianti} lists 23 transitions of Fe\,{\sc xi} in the EIS long wavelength channel (245--291\,\AA), most of these are either too weak to be detected or are blended with much stronger emission features from other species.
As a result, we only have measurements for 5 lines, which are summarised in Table 2. Portions of the EIS active region and limb spectra containing these lines are shown in Figure 1. In Table 10 we list the resultant observed line ratios (and their 1$\sigma$ errors), along with the theoretical results from the present calculations. Also given are the theoretical predictions from 
the {\sc chianti} V6.0 data base, which employs the electron excitation rates of Gupta \& Tayal (1999) and Bhatia \& 
Doschek (1996). The 266.39/257.55 and 266.60/257.55 ratios are relatively insensitive to changes in the electron density over the range N$_{e}$ = 10$^{9}$--10$^{11}$\,cm$^{-3}$, varying by only 13 and 8 per cent, respectively. However, 276.36/257.55 is N$_{e}$--sensitive, as may be seen in Figure 2. 
Hence we have determined values of N$_{e}$  for the active region and limb using the
200.03/202.04 and 203.17/202.04 line intensity ratios in Fe\,{\sc xiii}, which is  formed at a similar temperature to
Fe\,{\sc xi} in ionisation equilibrium (T$_{e}$ = 10$^{6.2}$\,K compared to 10$^{6.1}$\,K for Fe\,{\sc xi};
Bryans et al. 2009). For the active region, we measured 
200.03/202.04 = 0.31 and 203.17/202.04 = 0.16, both of which yield N$_{e}$ = 10$^{9.2}$\,cm$^{-3}$ from the 
calculations of Keenan et al. (2007). 
Similarly, for the limb spectrum the observed ratios are 200.03/202.04 = 0.47 and 203.17/202.04 = 0.24, 
which both indicate N$_{e}$ = 10$^{9.5}$\,cm$^{-3}$.
In Table 10 we therefore list the theoretical ratios 
generated at the temperature of maximum Fe\,{\sc xi} fractional abundance in ionisation equilibrium 
and at densities of 10$^{9.2}$ and 10$^{9.5}$\,cm$^{-3}$ for the active region and limb area, respectively.

\begin{table*}
\begin{minipage}{180mm}
  \caption{Fe\,{\sc xi} line ratios from the EIS spectra.}
  \begin{tabular}{lcccc}
  \hline
Feature & Line ratio &   Observed & Present & {\sc chianti} 
\\
& & & theory$^{a}$ & theory$^{a}$
\\
\hline
Active region & 257.77/257.55 & 0.50 $\pm$ 0.09 & 0.62 & 0.52
\\
Active region & 266.39/257.55 & 0.051 $\pm$ 0.011 & 0.12 & 0.083
\\
Active region & 266.60/257.55 & 0.32 $\pm$ 0.04 & 0.11 & 0.069
\\
Active region & 276.36/257.55 & 0.048 $\pm$ 0.024 & 0.032 & 0.038
\\
Limb & 257.77/257.55 & 0.48 $\pm$ 0.07 & 0.61 & 0.54
\\
Limb & 266.39/257.55 & 0.068 $\pm$ 0.013 & 0.12 & 0.084
\\
Limb & 266.60/257.55 & 0.16 $\pm$ 0.03 & 0.11 & 0.072
\\
Limb & 276.36/257.55 & 0.022 $\pm$ 0.009 & 0.041 & 0.051
\\
\hline
\end{tabular}

$^{a}$Present theoretical ratios and those from {\sc chianti} calculated at T$_{e}$ = 10$^{6.1}$\,K and
N$_{e}$ = 10$^{9.2}$\,cm$^{-3}$ (active region) or N$_{e}$ = 10$^{9.5}$\,cm$^{-3}$ (limb).

\end{minipage} 
\end{table*}

The 266.39\,\AA\ line was tentatively identified by Tr\"{a}bert (1998) in the solar flare spectra of Dere (1978), and the reasonable agreement between theory and observation for the 266.39/257.55 ratio in Table 10 (especially for the {\sc chianti} calculations) supports this identification. We note that {\sc chianti} lists the 
3s$^{2}$3p$^{4}$ $^{3}$P$_{1}$--3s$^{2}$3p$^{3}$($^{4}$S)3d $^{5}$D$_{1}$ transition at 273.41\,\AA, but no emission feature is detected at this wavelength in the EIS spectra. In addition, Brown et al. (2008) identify the 266.39\,\AA\ feature as an Fe\,{\sc xvii} transition. However, the fact that the measured 266.39/257.55 ratios are actually somewhat smaller than theory indicates that Fe\,{\sc xvii} probably makes a negligible contribution to the 266.39\,\AA\ line intensity.

Tr\"{a}bert (1998) also tentatively identified the 266.60\,\AA\ line of Fe\,{\sc xi} in the Dere (1978) spectra, but in this case the experimental 266.60/257.55 ratios in Table 10 are larger than the theoretical values, at least for the active region. This suggests that the 266.60\,\AA\ feature is blended, which is supported by the relatively large width found for the line. However, an inspection of {\sc chianti} and other line lists reveals no likely 
blending candidates. We note that {\sc chianti} lists a wavelength of 273.57\,\AA\ for the 
3s$^{2}$3p$^{4}$ $^{3}$P$_{1}$--3s$^{2}$3p$^{3}$($^{4}$S)3d $^{5}$D$_{0}$ transition, which unfortunately would be blended with Fe\,{\sc xiv} 273.55\,\AA. However, the experimental Fe\,{\sc xiv} 273.55/274.20 ratios in the EIS spectra are in good agreement with the {\sc chianti} predictions, indicating that Fe\,{\sc xi} does not make a contribution to the 273.55\,\AA\ line intensity. We also note that Brown et al. (2008) detect the 266.60\,\AA\ 
line, but no identification is provided.

\begin{figure}
\epsfig{file=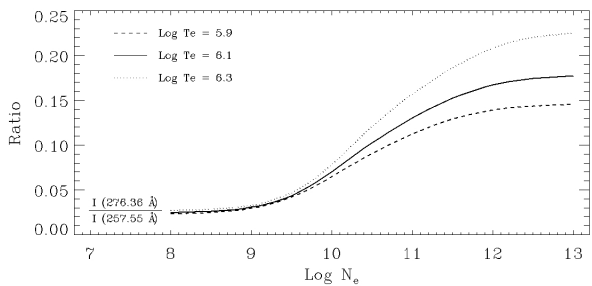,angle=0,width=8.5cm}
\caption{The theoretical Fe\,{\sc xi}
emission-line intensity ratio
I(276.36\,\AA)/I(257.55\,\AA), where I is in energy units,
plotted as a function of logarithmic electron density
(N$_{e}$ in cm$^{-3}$) at the temperature
of maximum Fe\,{\sc xi} fractional abundance in ionization
equilibrium, T$_{e}$ = 10$^{6.1}$\,K (Bryans et al.
2009), plus $\pm$0.2 dex about this
value. }
\end{figure}

Of particular interest is the 
276.36\,\AA\ line, which was unidentified in solar observations until recently detected by 
Brown et al. (2008) in the EIS limb spectrum.
However, these authors identify the feature as Fe\,{\sc xii} rather than Fe\,{\sc xi}, 
although this is almost certainly a typographical error as the transition is correctly given as 
3s$^{2}$3p$^{4}$ $^{3}$P$_{2}$--3s3p$^{5}$ $^{1}$P$_{1}$.  (We  note in passing that
Brown et al. incorrectly list the 257.77\,\AA\ line as having a $^{5}$D$_{4}$
upper level, when it should be $^{5}$D$_{2}$). 
An inspection of Table 10 reveals good agreement between theory and observation for 276.36/257.55, confirming the
identification of the 276.36\,\AA\ feature as the 
3s$^{2}$3p$^{4}$ $^{3}$P$_{2}$--3s3p$^{5}$ $^{1}$P$_{1}$ line of Fe\,{\sc xi}. However, further observations of the transition would be useful, as the 276.36/257.55 
ratio potentially provides a useful EIS long wavelength channel
density diagnostic for the Fe\,{\sc xi}--emitting region of a plasma, as it varies by a factor of 
6.8 between N$_{e}$ = 10$^{8}$ and 10$^{12}$\,cm$^{-3}$.

In Table 10 we also list the observed and theoretical 257.77/257.55 ratios, for which there is good agreement, 
indicating that both lines are well resolved in the EIS spectra and free from blends. To check this we 
have generated synthetic active region, flare and quiet Sun spectra 
using V6.0 of the {\sc chianti} data base. These reveal
no significant blends for either the 257.55 or 257.77\,\AA\ transitions, nor indeed for the 276.36\,\AA\ feature.
We note that Jup\'{e}n et al. (1993) originally identified the 257.55 and 257.77\,\AA\ features as being the (2--3) and (2--2) components, respectively, of the 
3s$^{2}$3p$^{4}$ $^{3}$P$_{J}$--3s$^{2}$3p$^{3}$($^{4}$S)3d $^{5}$D$_{J^\prime}$
multiplet. Subsequently, Tr\"{a}bert (1998) reclassified the lines as the (2--2) and (2--1) components, with (2--3) lying at 257.26\,\AA\ and hence blended with a strong Fe\,{\sc x} line. However, the theoretical (2--1)/(2--2) ratio is 0.31 (present calculations) or 0.27 ({\sc chianti}), much smaller than the experimental values of around 0.50 for 257.77/257.55.  Hence the EIS observations do not support the Tr\"{a}bert classifications, but rather those of Jup\'{e}n et al. A definitive answer would be provided by the unambiguous detection of the (1--2) or (2--1) components of the multiplet. 
Unfortunately, the (1--2) transition is predicted to be too weak to observe, with a theoretical intensity of less than 0.1 per cent that of the 257.55\,\AA\ feature. Similarly, the (2--1) transition is predicted to lie at 264.92\,\AA\ by {\sc chianti}, and cannot be observed as it lies between lines of Fe\,{\sc xiv} and Fe\,{\sc xvi}. 

\subsection{SERTS observations}

The SERTS first-order wavelength range ($\sim$\,235--449\,\AA) contains numerous Fe\,{\sc xi} emission features,
resulting in many possible line ratio combinations. Following 
Keenan et al. (2007), we categorise 3 types of ratio, namely:
\\
\\
(i) branching ratios,  where the transitions arise from a common upper level. In this instance
the ratio is predicted to be constant under most plasma conditions, an exception being
when significant opacity is present, which should not apply to the Fe\,{\sc xi}
lines considered here;

(ii) ratios which are predicted to be relatively insensitive to changes 
in T$_{e}$ and N$_{e}$ over the range of plasma parameters of interest;

(iii)
ratios which are predicted to vary significantly with N$_{e}$, and hence may 
provide useful electron density 
diagnostics.
\\
\\
Ratios in categories (i) and (ii) are the most useful for 
identifying and assessing the importance of blends, as well as investigating possible
errors in the adopted
atomic data, 
as one does not need to accurately know the plasma parameters 
to generate a theoretical ratio for comparison with observations.
In Tables 11 and 12 we therefore list the 
observed line ratios for the SERTS data sets  which fall into
category (i) or (ii), respectively (along with the associated 1$\sigma$ errors),  as well as
the theoretical values both from the 
present calculations
and {\sc chianti}. 
Once again, following Keenan et al. (2007), we have defined 
`relatively insensitive' as being those ratios which
are predicted to vary by less than $\pm$20 per cent when the electron density 
is changed by a factor of 2 (i.e. $\pm$0.3 dex). In fact, most of the ratios in
Table 12 show much lower sensitivity to variations in N$_{e}$.
For example, at T$_{e}$ = 10$^{6.1}$\,K, the 341.11/352.67 ratio only changes from 0.27 
at  N$_{e}$ = 10$^{9}$\,cm$^{-3}$to 0.31 at N$_{e}$ = 10$^{10}$\,cm$^{-3}$, while 
257.55/352.67 varies from 0.13 to 0.16 over the same density interval.
 As noted by Keenan et al., 
most of the electron densities derived for 
SERTS active regions from species formed 
at similar temperatures to Fe~{\sc xi}
are consistent with log N$_{e}$ = 9.4$\pm$0.3. This is also the case for both quiet Sun regions
and the off-limb area, where diagnostics indicate log N$_{e}$ = 9.1$\pm$0.3 (Brosius et al. 1996).
The theoretical results in Table 12 have therefore been calculated at the temperature of
maximum Fe\,{\sc xi} fractional abundance in ionization equilibrium, 
T$_{e}$ = 10$^{6.1}$\,K (Bryans et al. 2009), and at  N$_{e}$ = 10$^{9.4}$\,cm$^{-3}$ (for the active regions)
or N$_{e}$ = 10$^{9.1}$\,cm$^{-3}$ (quiet Sun and off-limb). 
However, we note that
changing the 
adopted value of T$_{e}$ 
by $\pm$0.2 dex or the density by $\pm$0.5 dex does not significantly alter 
the discussions below.

\begin{table*}
\begin{minipage}{180mm}
  \caption{Comparison of theory and observation for SERTS line ratios involving transitions from common upper 
levels.}
  \begin{tabular}{lcccc}
  \hline
Feature & Line ratio &   Observed & Present & {\sc chianti} 
\\
& & & theory & theory
\\
\hline
SERTS89--AR & 341.11/356.53 & 2.1 $\pm$ 0.6 & 2.0 & 2.0
\\
SERTS91--QS & 341.11/356.53 & 3.4 $\pm$ 1.7 & 2.0 & 2.0
\\
SERTS93--AR & 341.11/356.53 & 3.7 $\pm$ 1.7 & 2.0 & 2.0
\\
SERTS89--AR & 341.11/358.67 & 0.52 $\pm$ 0.10 & 1.6 & 1.6
\\
SERTS91--AR & 341.11/358.67 & 1.0 $\pm$ 0.3 & 1.6 & 1.6
\\
SERTS93--QS & 341.11/358.67 & 1.3 $\pm$ 0.5 & 1.6 & 1.6
\\
SERTS89--AR & 358.67/356.53 & 4.0 $\pm$ 1.0 & 1.2 & 1.3
\\
SERTS91--QS & 358.67/356.53 & 0.83 $\pm$ 0.60 & 1.2 & 1.3
\\
SERTS93--AR & 358.67/356.53 & 3.2 $\pm$ 1.5 & 1.2 & 1.3
\\
SERTS89--AR & 369.16/352.67 & 0.29 $\pm$ 0.05 & 0.30 & 0.30
\\
SERTS91--AR & 369.16/352.67 & 0.36 $\pm$ 0.11 & 0.30 & 0.30
\\
SERTS91--QS & 369.16/352.67 & 0.39 $\pm$ 0.12 & 0.30 & 0.30
\\
SERTS91--OL & 369.16/352.67 & 0.25 $\pm$ 0.05 & 0.30 & 0.30
\\
SERTS93--AR & 369.16/352.67 & 0.34 $\pm$ 0.09 & 0.30 & 0.30
\\
SERTS93--QS & 369.16/352.67 & 0.27 $\pm$ 0.08 & 0.30 & 0.30
\\
SERTS89--AR & 406.79/352.67 & 0.022 $\pm$ 0.009 & 0.023 & 0.027
\\
\hline
\end{tabular}
\end{minipage} 
\end{table*}

\begin{table*}
\begin{minipage}{180mm}
  \caption{Comparison of theory and observation for  SERTS line ratios which are only weakly 
N$_{e}$--dependent.}
  \begin{tabular}{lcccc}
  \hline
Feature & Line ratio &   Observed & Present & {\sc chianti} 
\\
& & & theory$^{a}$ & theory$^{a}$
\\
\hline
SERTS89--AR & 257.55/352.67 & 0.22 $\pm$ 0.10 & 0.14 & 0.19
\\
SERTS89--AR & 341.11/352.67 & 0.29 $\pm$ 0.05 & 0.29 & 0.28
\\
SERTS91--AR & 341.11/352.67 & 0.55 $\pm$ 0.13 & 0.29 & 0.28
\\
SERTS91--QS & 341.11/352.67 & 0.52 $\pm$ 0.12 & 0.28 & 0.27
\\
SERTS91--OL & 341.11/352.67 & 0.41 $\pm$ 0.08 & 0.28 & 0.27
\\
SERTS93--AR & 341.11/352.67 & 0.40 $\pm$ 0.07 & 0.29 & 0.28
\\
SERTS93--QS & 341.11/352.67 & 0.28 $\pm$ 0.07 & 0.28 & 0.27
\\
SERTS97--AR & 341.11/352.67 & 0.28 $\pm$ 0.07 & 0.29 & 0.28
\\
SERTS89--AR & 349.04/308.54 & 0.25 $\pm$ 0.12 & 0.65 & 0.50
\\
SERTS97--AR & 349.04/308.54 & 0.37 $\pm$ 0.16 & 0.65 & 0.50
\\
SERTS89--AR & 356.53/352.67 & 0.14 $\pm$ 0.03 & 0.14 & 0.14
\\
SERTS91--QS & 356.53/352.67 & 0.15 $\pm$ 0.07 & 0.14 & 0.13
\\
SERTS93--AR & 356.53/352.67 & 0.11 $\pm$ 0.05 & 0.14 & 0.14
\\
SERTS89--AR & 358.67/352.67 & 0.55 $\pm$ 0.10 & 0.18 & 0.18
\\
SERTS91--AR & 358.67/352.67 & 0.53 $\pm$ 0.16 & 0.18 & 0.18
\\
SERTS91--QS & 358.67/352.67 & 0.13 $\pm$ 0.08 & 0.17 & 0.17
\\
SERTS91--OL & 358.67/352.67 & 0.15 $\pm$ 0.08 & 0.17 & 0.17
\\
SERTS93--AR & 358.67/352.67 & 0.34 $\pm$ 0.08 & 0.18 & 0.18
\\
SERTS93--QS & 358.67/352.67 & 0.21 $\pm$ 0.07 & 0.17 & 0.17
\\
\hline
\end{tabular}

$^{a}$Present theoretical ratios and those from {\sc chianti} calculated at T$_{e}$ = 10$^{6.1}$\,K and
N$_{e}$ = 10$^{9.4}$\,cm$^{-3}$ (for active regions) or N$_{e}$ = 10$^{9.1}$\,cm$^{-3}$ (quiet Sun and off-limb regions).

\end{minipage} 
\end{table*}

An inspection of Table 12 reveals good agreement between theory and observation for the 257.55/352.67 ratio, providing support for our identification of the 257.55\,\AA\ line. This is the first time this feature has been detected in SERTS observations, although it has previously been identified by Jup\'{e}n et al. (1993) in the solar spectrum of Behring et al. (1976), as well as in EIS data (see Section 4.1).

The measured 349.04/308.54 ratio in Table 12 for the SERTS89 active region is smaller than the present theoretical result or that from {\sc chianti}, which indicates blending in the 308.54\,\AA\ line. Brosius et al. (1998) note that the Fe\,{\sc xi} 308.54\,\AA\ feature is blended with an Fe\,{\sc vi} transition, but from the {\sc chianti} synthetic spectra the latter is predicted to contribute less than 0.1 per cent of the total intensity. Indeed, {\sc chianti} indicates that there are lines of Ni\,{\sc xiii}, Al\,{\sc vi} and Ne\,{\sc iii} which make a larger contribution to the blend than Fe\,{\sc vi}, but even then the summed intensity of these 3 transitions is only 5 per cent that of Fe\,{\sc xi}. On the other hand, the 308.54\,\AA\ line width in the SERTS89 active region spectrum is much larger than in the other SERTS data sets, as may be seen from Figure 3 and Tables 3--9. Furthermore, the feature lies at 308.58\,\AA\ in the SERTS89 observations, compared to 308.54\,\AA\ in the remaining SERTS data sets and other solar spectra, such as those from the S082A instrument on board {\em Skylab} (Dere 1978). Both the large line width and wavelength shift are symptomatic of blending, but if such a species is present it is difficult to explain why it should affect one active region and not others. A more likely explanation is a fault with the SERTS spectrum, although an inspection of the film reveals no apparent flaws.

The experimental 349.04/308.54 ratio for the SERTS97 active region is in good agreement with theory, indicating that both these features are reliably observed and free from blends. 
Thomas \& Neupert (1994) point out that the Fe\,{\sc xi} 349.04\,\AA\ line in the SERTS89 data is blended with the second-order Fe\,{\sc x} 174.52\,\AA\ transition, which appears at 349.04\,\AA\ in first-order. However, as noted in Section 3, the version of SERTS flown in 1997 incorporated a multilayered-coated grating, which enhanced the instrument sensitivity in first-order compared to the SERTS89 observations, hence removing the effect of the blend with Fe\,{\sc x} 174.52\,\AA.
On the other hand, this blending is evident in the spectrum from the SERTS flight in 1995, where the instrument sensitivity was enhanced in the second-order waveband (Brosius et al. 1998), and the observed 349.04/308.54 ratio is 3.4, much larger than the theoretical value of about 0.6.

\begin{figure}
\epsfig{file=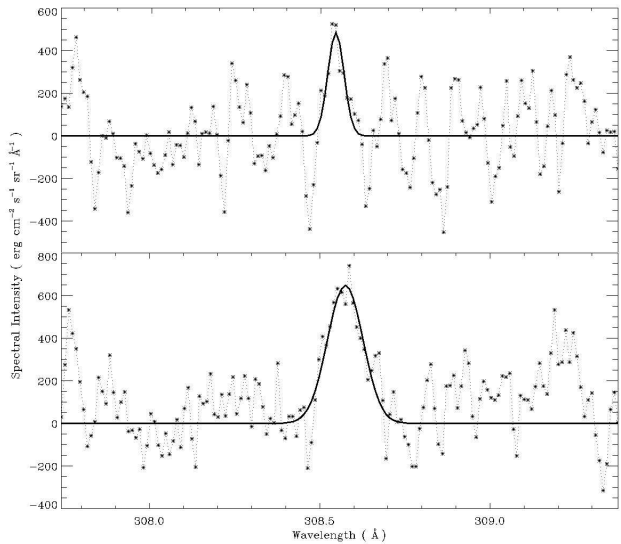,angle=0,width=8.5cm}
\caption{Plot of the SERTS 1989 (lower panel) and 1993 (upper panel) active region spectra in the
307.8--309.3\,\AA\ wavelength range.
The profile fits to the Fe\,{\sc xi} 308.54\,\AA\ feature
are shown by solid lines.}
\end{figure}

Both the experimental 341.11/356.53 and 341.11/352.67 ratios in Tables 11 and 12 are generally larger than the theoretical values, which {\sc chianti} indicates is due to blending of 341.11\,\AA\ with an Fe\,{\sc ix} transition. The latter is predicted to contribute about 33 per cent to the total 341.11\,\AA\ line intensity, which is consistent with the observations. For example, for the SERTS93 active region, reducing the intensity of the 341.11\,\AA\ line by 33 per cent leads to a revised 341.11/352.67 ratio of 0.26, in good agreement with the theoretical value of 0.29. 

The measured 356.53/352.67 ratios in Table 12 are all in good agreement with theory, indicating that both the 352.67 and 356.53\,\AA\ lines are well observed and free from blends. Similarly, there are no discrepancies between theory and observation for 369.16/352.67, and indeed the average of the 6 measurements in Table 11 is 0.32 $\pm$ 0.05, which compares very well with the theoretical result of 0.30, showing that the 369.16\,\AA\ transition is also free from blends. We note that the 352.67, 356.53 and 369.16\,\AA\ lines of Fe\,{\sc xi} have also been recently observed by the EUNIS solar instrument (Brosius et al. 2008), and show good agreement between theory and experiment, with measured ratios of 356.53/352.67 = 0.16 $\pm$ 0.03 and 369.16/352.67 = 0.30 $\pm$ 0.05, compared to the theoretical estimates of 0.14 and 0.30, respectively.

Young et al. (1998) note that the 358.67\,\AA\ line of Fe\,{\sc xi} is blended with several transitions of Si\,{\sc xi}, Ne\,{\sc iv} and Fe\,{\sc xiv}. However, the observed 358.67/356.53 and 358.67/352.67 ratios in Tables 11 and 12 are only much larger than the theoretical values for the active region spectra, with measurements for the quiet Sun and off-limb regions being in good agreement with theory. This implies that blending of 358.67\,\AA\ is only significant in active regions. By contrast, the synthetic active region and quiet Sun spectra generated with {\sc chianti} predict that Fe\,{\sc xi} should make a smaller contribution to the total 358.67\,\AA\ line intensity in the latter (less than 20 per cent compared to about 35 per cent in an active region), and hence that blending of the Fe\,{\sc xi} feature should in fact be more severe in quiet Sun spectra. However, Fe\,{\sc xi} observations from other solar missions support the present findings. For example, for a quiet solar region observed with the CDS instrument on the SOHO satellite, Landi et al. (2002) measured Fe\,{\sc xi} ratios of 358.67/356.53 = 1.1 $\pm$ 0.1 and 
358.67/352.67 = 0.14 $\pm$ 0.02, in excellent agreement with the theoretical values of 1.2 and 0.17, respectively.
Better agreement between theory and observation for the quiet Sun compared to active regions is indicative of blending with lines formed at higher temperatures than Fe\,{\sc xi}. Our results would therefore suggest that Si\,{\sc xi} and Fe\,{\sc xiv} (both of which are formed at slightly higher temperatures than Fe\,{\sc xi}) make a larger contribution to the 358.67\,\AA\ line intensity than currently predicted by {\sc chianti}, while Ne\,{\sc iv} makes a smaller contribution.

Finally, the observed 406.79/352.67 ratio in Table 11  is in very good agreement with the present calculations and also those from {\sc chianti}, supporting the identification of the 406.79\,\AA\ line in the SERTS89 active region spectrum by Brickhouse et al. (1995).

\subsection{Electron density diagnostics from SERTS spectra}

In Figures 4 and 5 we plot the electron density diagnostics 308.54/352.67 and 349.04/352.67, respectively, while in Table 13 we list the observed ratios (and the associated 1$\sigma$ errors) along with the values of N$_{e}$ derived using the theoretical results  at the temperature of maximum Fe\,{\sc xi} fractional abundance in ionisation equilibrium, T$_{e}$ = 10$^{6.1}$\,K (Bryans et al. 2009). 
However, we note that changing T$_{e}$ by $\pm$0.2 dex
would lead to a variation in the derived
values of N$_{e}$ of at most $\pm$0.1 dex.
Also listed in the table is the factor by which the relevant ratio is predicted to vary  
between N$_{e}$ = 10$^{8}$
and 10$^{11}$\,cm$^{-3}$.

\begin{table*}
\begin{minipage}{180mm}
  \caption{Electron density diagnostic line ratios from the SERTS spectra.}
  \begin{tabular}{lcccc}
  \hline
Feature & Line ratio &   Observed & log N$_{e}$$^{a}$ & Ratio variation$^{b}$ 
\\
\hline
SERTS89--AR & 308.54/352.67 & 0.66 $\pm$ 0.15 & 10.7$^{+0.5}_{-0.3}$ & 8.4
\\
SERTS91--AR & 308.54/352.67 & 0.16 $\pm$ 0.05 & 9.4$^{+0.2}_{-0.5}$ & 8.4
\\
SERTS91--QS & 308.54/352.67 & 0.17 $\pm$ 0.06 & 9.4$^{+0.3}_{-0.5}$ & 8.4
\\
SERTS91--OL & 308.54/352.67 & 0.18 $\pm$ 0.06 & 9.5$^{+0.2}_{-0.5}$ & 8.4
\\
SERTS93--AR & 308.54/352.67 & 0.16 $\pm$ 0.07 & 9.4$^{+0.3}_{-1.3}$ & 8.4
\\
SERTS93--QS & 308.54/352.67 & 0.14 $\pm$ 0.07 & 9.2$^{+0.4}_{-\infty}$ & 8.4
\\
SERTS97--AR & 308.54/352.67 & 0.25 $\pm$ 0.08 & 9.8$^{+0.2}_{-0.4}$ & 8.4
\\
SERTS89--AR & 349.04/352.67 & 0.16 $\pm$ 0.08 & 9.9$^{+\infty}_{-0.7}$ & 14.7
\\
SERTS97--AR & 349.04/352.67 & 0.092 $\pm$ 0.034 & 9.3$^{+0.3}_{-0.4}$ & 14.7
\\
\hline
\end{tabular}

$^{a}$Determined from present line ratio calculations at T$_{e}$ = 10$^{6.1}$\,K; N$_{e}$ in cm$^{-3}$. 
\\
$^{b}$Factor by which the theoretical line 
ratio varies between N$_{e}$ = 10$^{8}$ and 10$^{11}$\,cm$^{-3}$.
\end{minipage} 
\end{table*}

\begin{figure}
\epsfig{file=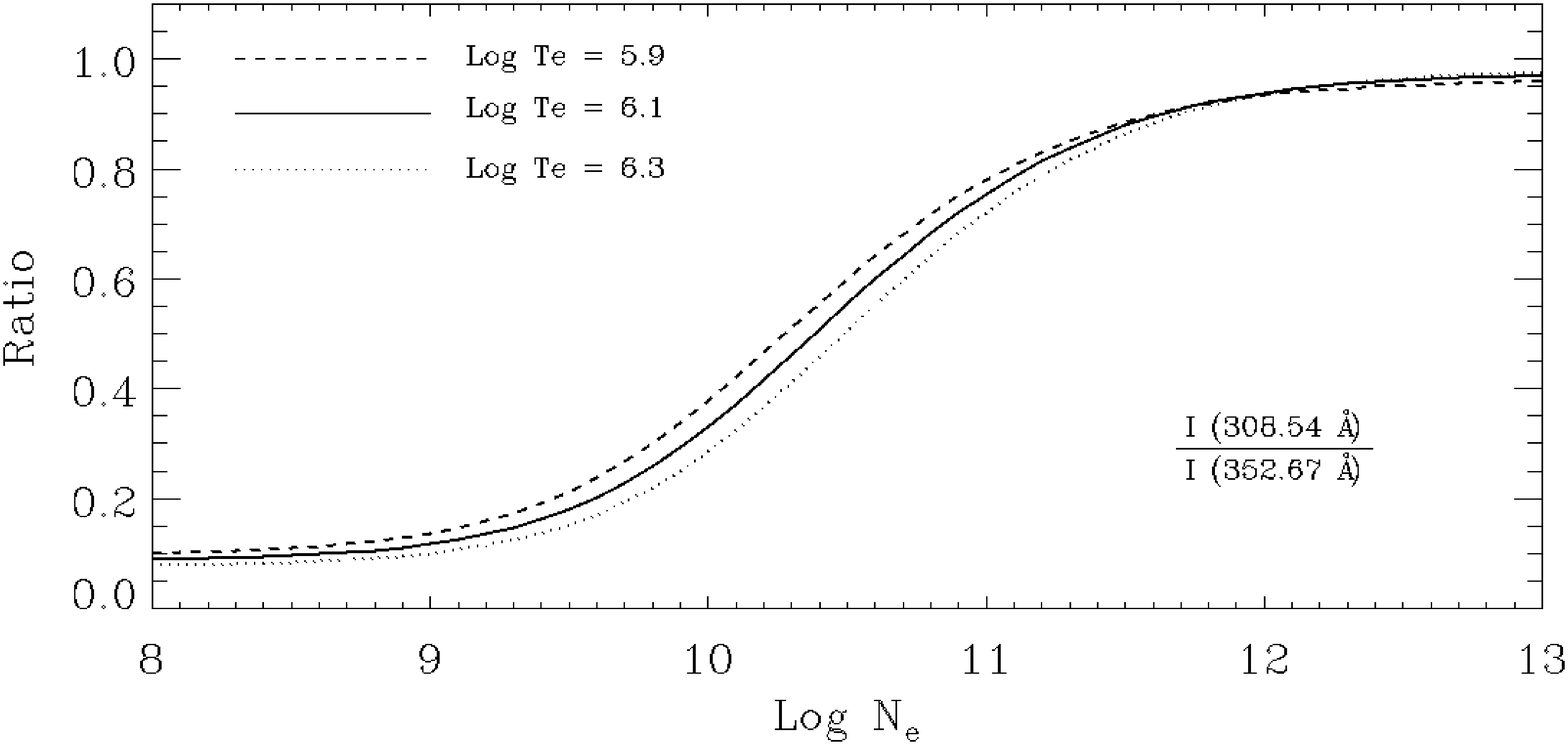,angle=0,width=8.5cm}
\caption{The theoretical Fe\,{\sc xi}
emission-line intensity ratio
I(308.54\,\AA)/I(352.67\,\AA), where I is in energy units,
plotted as a function of logarithmic electron density
(N$_{e}$ in cm$^{-3}$) at the temperature
of maximum Fe\,{\sc xi} fractional abundance in ionization
equilibrium, T$_{e}$ = 10$^{6.1}$\,K (Bryans et al.
2009), plus $\pm$0.2 dex about this
value. }
\end{figure}

\begin{figure}
\epsfig{file=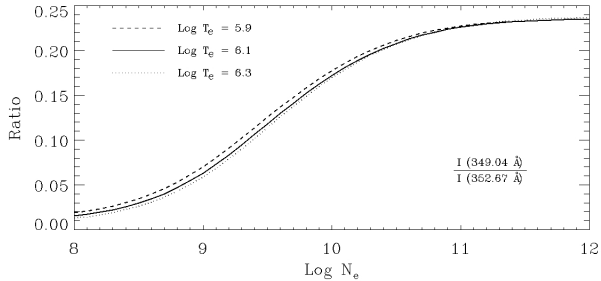,angle=0,width=8.5cm}
\caption{The theoretical Fe\,{\sc xi}
emission-line intensity ratio
I(349.04\,\AA)/I(352.67\,\AA), where I is in energy units,
plotted as a function of logarithmic electron density
(N$_{e}$ in cm$^{-3}$) at the temperature
of maximum Fe\,{\sc xi} fractional abundance in ionization
equilibrium, T$_{e}$ = 10$^{6.1}$\,K (Bryans et al.
2009), plus $\pm$0.2 dex about this
value. }
\end{figure}

An inspection of Table 13 reveals that the densities derived from 308.54/352.67 are generally in very good agreement with those estimated for the SERTS solar features from line ratios in species formed at similar temperatures to Fe\,{\sc xi}, which as noted in Section 4.2  average about log N$_{e}$ = 9.4 $\pm$ 0.3 for active regions and log N$_{e}$ = 9.1 $\pm$ 0.3 for quiet Sun and off-limb areas. Hence we can state that, in general, the 308.54/352.67 ratio appears to provide a good N$_{e}$--diagnostic. The exception is the SERTS89 active region, where the derived density is more than an order of magnitude larger than expected. However, as already discussed in Section 4.2, there are problems with the 308.54\,\AA\ feature in the SERTS89 active region observations which make its measurement uncertain.

The electron density derived for the SERTS97 active region from the 349.04/352.67 ratio is in good agreement 
with those found from other species, but in the case of the SERTS89 active region the density is much higher than expected. However, as noted by Thomas \& Neupert (1994), the 349.04\,\AA\ line in the SERTS89  observations is blended with the second-order Fe\,{\sc x} 174.52\,\AA\ feature.
We can estimate its contribution using the nearby Fe\,{\sc x} 177.24\,\AA\ transition, which appears in first-order at 354.48\,\AA. The first-order intensity of this line is 3.6
erg\,cm$^{-2}$\,s$^{-1}$\,sr$^{-1}$, and the 174.52\,\AA\ feature is predicted to be a factor of 1.8 stronger (Keenan et al. 2008), yielding an intensity of 6.5 erg\,cm$^{-2}$\,s$^{-1}$\,sr$^{-1}$. This in turn gives a corrected intensity for Fe\,{\sc xi} 349.04\,\AA\ of 19.7 erg\,cm$^{-2}$\,s$^{-1}$\,sr$^{-1}$, and revised experimental 349.04/352.67 ratio of 0.12. From Figure 5, this implies N$_{e}$ = 10$^{9.5}$\,cm$^{-3}$, in good agreement with other diagnostics.

Although the 308.54/352.67 ratio is probably the most useful Fe\,{\sc xi} density diagnostic currently available in the 257--407\,\AA\ wavelength range, 349.04/352.67 is superior in many respects. 
The lines are closer together, reducing the effects of possible errors in the instrument intensity calibration, and the ratio varies by a larger factor over the density range N$_{e}$ = 10$^{8}$--10$^{11}$\,cm$^{-3}$
(14.7 c.f. 8.4 for 308.54/352.67).
However, the 349.04\,\AA\ feature is relatively weak, and is also blended with Fe\,{\sc x} 174.52\,\AA, unless 
the first-order instrument response is enhanced, as for the SERTS97 observations.

\section*{Acknowledgments}

DBJ and KMA acknowledge financial support from STFC and EPSRC, respectively.
ROM acknowledges support from the NASA Postdoctoral Program at the Goddard Space Flight Center, administered by Oak Ridge Associated Universities through a contract with NASA. {\em Hinode} is a Japanese mission developed and launched by ISAS/JAXA, collaborating with NAOJ as a domestic partner, NASA and STFC as international partners. Scientific operation of the {\em Hinode} mission is conducted by the {\em Hinode} science team organised at ISAS/JAXA. Support for the postlaunch operation is provided by JAXA and NAOJ, STFC, NASA, ESA and NSC (Norway).
The SERTS rocket programme is
supported by RTOP grants from the Solar Physics Office   
of NASA's Space Physics Division.                        
JWB acknowledges additional NASA support under
grant NAG5--13321.                                       
FPK is grateful to AWE Aldermaston for the award of a William Penney
Fellowship. The authors thank Peter van Hoof for the use of his
Atomic Line List. {\sc chianti} is a collaborative project 
involving the Naval Research Laboratory (USA), Rutherford
Appleton Laboratory (UK), and the Universities of Florence 
(Italy) and Cambridge (UK). We are very grateful to the referee, Peter Young, for his comments on an earlier version of the paper, in particular regarding the analysis of the EIS spectra.

\bsp

\label{lastpage}

\end{document}